# Lattice Kerker effect in the hexagonal boron nitride antenna array

Viktoriia E. Babicheva

*College of Optical Sciences, University of Arizona, Tucson, AZ, USA*

vbab.dtu@gmail.com

**ABSTRACT.** Subwavelength particles with hyperbolic light dispersion in the constituent medium are a promising alternative to plasmonic, high-refractive-index dielectric, and semiconductor structures in the practical realization of nanoscale optical elements. Hexagonal boron nitride (hBN) is a layered van der Waals material with natural hyperbolic properties and low-loss phonon-polaritons at the same time. In this work, we consider multipole excitations and antennas properties of hBN particles with an emphasis on the periodic arrangement and collective array modes. We analyze excitation of lattice resonances in the antenna array and effect of resonance shifts and overlap with other multipoles supported by particles in the lattice. In such periodic structure, a decrease of reflectance from the array is achieved with appropriate lattice spacing (periods) where the electric and magnetic multipoles overlap, and the resonance oscillations are in phase and comparable in magnitude. We theoretical demonstrate that in this case, generalized Kerker condition is satisfied, and hBN antennas in the array efficiently scatter light in the predominantly forward direction resulting in near-zero reflectance. The resonant lattice Kerker effect with hyperbolic-medium antennas can be applied in developing metasurfaces based on hBN resonators for mid-infrared photonics.

## INTRODUCTION

Antennas made of a natural hyperbolic material (hexagonal boron nitride) have been proposed to be used in designing ultra-thin optical elements [1-3]. For now, nanoscale metal-dielectric structures have been considered for a wide range of applications such as antennas and cavities with enhanced light-matter interaction, photovoltaics and light harvesting, high-resolution optical microscopy, and others [4-11]. Semiconductors and high-refractive-index dielectrics have been shown a promising alternative to plasmonic structures [12-20]. The interplay of electric and magnetic resonances of the particles can result in directional scattering from a single particle, suppression or increase of reflection from the array of such particles, and Kerker effect with the generalized condition of scattering compensation satisfied [21-27]. Recently, it has been shown that hyperbolic-dispersion medium is a viable counterpart to metal-dielectric structures for realizing subwavelength antennas and scattering building blocks in metasurfaces [2,3].

The permittivity tensor components of the hyperbolic medium have different signs causing hyperbola-like dispersion of propagating eigenmodes [28,29]. In case of low losses, this medium supports waves with extremely high propagation constants (far exceeding those in plasmonic structures), resulting in the enhanced spontaneous emission, possibility of anomalous heat transfer [30-35], subwavelength mode confinement in waveguides [36-40] and tapers [41,42], etc. Similar to localized surface plasmon resonances or boundary-reflected modes in high-refractive-index nanoparticles, one can achieve a resonance of highly confined waves in the particles with hyperbolic medium and deeply subwavelength dimensions.

Collective lattice resonances in periodic arrays have been studied for both plasmonic and all-dielectric particles, and sharp Fano-like features have been analyzed in absorption, reflection, and transmittance profiles of various structures [43-48,51]. Previously, we have theoretically demonstrated that particle made of the hyperbolic medium can possess resonances at the wavelength defined by the antennas dimension and period of the structure [2,3]. In particular, hexagonal boron nitride (hBN) antennas have been proposed as subwavelength resonators, and the possibility of multipole resonance excitations have been pointed out with an emphasis on scattering properties of particles. Here we show that the periodic arrangement of the hBN antennas results in pronounced lattice resonances that are tunable by the corresponding period of the structure and the resonant overlaps of the hBN cuboid multipoles cause directional scattering and resonant lattice Kerker effect.

## RESULTS

In this work, we consider an array of hBN antennas that have a cuboid shape with dimensions $a_x$ = 0.9 µm, $a_y$ = 0.5 µm, and $a_z$ = 2 µm (Fig. 1). For the material with hyperbolic dispersion, we study van der Waals material hBN that has Type II hyperbolic dispersions in the range 6.2 – 7.4 µm. Material permittivity parameters are taken from the work [49]. We perform numerical simulations with CST Microwave Studio frequency-domain solver based on finite element method. One antenna is considered in the unit cell, periodic boundary conditions are applied in $x$- and $y$-directions, and the artificial perfectly matched layers are considered in the $z$-direction. Antennas are assumed in a uniform medium with refractive index $n$ = 1, i.e. there is neither substrate nor superstrate.

Bulk hBN medium support propagation of high-$k$ waves, and the cuboid antenna has resonances stemming from multiple reflections of these high-$k$ waves on cuboid boundaries. These resonances are similar to Mie resonances in dielectric particles and Fabry-Perot resonances in general. In contrast to surface plasmon modes, the hBN resonances studied here have maximum field localization inside the particles and not on its surface [3]. As has been shown for the hBN resonators, the higher-order multipole resonances appear at the larger wavelength with respect to dipole resonances (anomalous scaling law [50]), and one can find a detailed analysis of the hBN modes in the cuboid resonator in the earlier works [2,3].

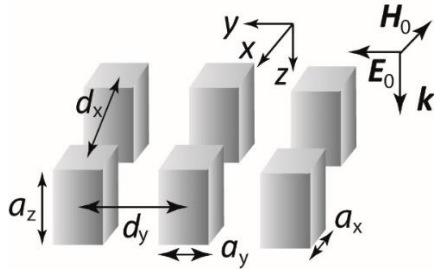

Figure 1. Schematic view of the hBN antenna array with periods $d_x$ and $d_y$ along corresponding directions. Antenna dimensions are $a_x$, $a_y$, and $a_z$, and electric field $E$ of the incident light is along the $y$-direction.

We consider the periodic arrangement of hBN antennas in the array where period $d_x$ = 2.5 µm, and $d_y$ is varied within the range 6.2 – 7.3 µm. Electric field $E$ of the incident light is along the $y$-axis. Rayleigh anomaly (corresponding to $d_y$) approaches multipole resonances of hBN antennas for $d_y$ = 6.7 – 7.2 µm and strongly affects mode profiles. Dipole and quadrupole resonances are more sensitive to the period along the multipole emission axis. In our case, for polarization along the y-axis and varied period $d_y$, only magnetic dipole and electric quadrupole resonances effectively couple to the lattice and experience resonant lattice effect for $\lambda \approx d_y$ [44,47,48]. At the same time, the electric dipole and magnetic quadrupole resonances do not have lattice resonance at $\lambda \approx d_y$.

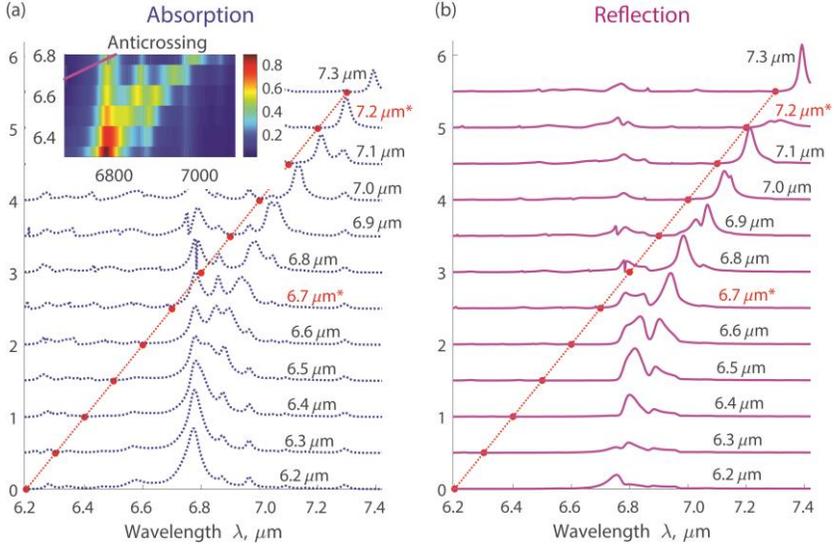

Figure 2. (a) Absorption and (b) reflection from the hBN antenna array. The resonator dimensions are $a_x = 0.9$ μm, $a_y = 0.5$ μm, and $a_z = 2$ μm; array period in the x-direction is $d_x = 2.5$ μm; and the electric field $E$ of the incident light is along the y-direction. Different lines correspond to different periods of the array within the range $d_y = 6.2 - 7.3$ μm, and the red dotted lines denote wavelength of Rayleigh anomaly. Lines for the periods $d_y = 6.7$ and 7.2 μm are marked in color, and the corresponding spectra are shown below in Fig. 3. On the ordinate axis, each line of the plot is offset by 0.5 with respect to the previous one. Absorption profiles show multiple antennas resonances, spectral positions of some resonances change following the Rayleigh anomaly, and upon resonance overlap, reflection is suppressed. Inset: mode anticrossing which indicates their coupling observed for the periods $d_y = 6.5 - 6.8$ μm.

From the calculations in Figure 2, one can see the well-pronounced lattice resonance following Rayleigh anomaly indicating a response from magnetic dipole and electric quadrupole moment. For the periods $d_y = 6.5 - 6.8$ μm, one can observe mode anticrossing (inset in Fig. 2a), which means modes are not independent but rather coupled and affect each other excitations. At the same time, some modes remain with unchanged position and are not affected by the lattice. Upon an overlap of the lattice resonances with unaffected antenna resonances, one can observe a decrease of reflection which is generalized lattice Kerker effect.

To emphasize the effect and analyze an origin of the reflection suppression, we consider several periods of the structure where the decrease of reflection is the most pronounced: $d_y = 6.7$ μm (Fig. 3a) and $d_y = 7.2$ μm (Fig. 3b). One can clearly see that for $d_y = 7.2$ μm, lattice resonance overlaps with the unaffected antennas resonances at $\lambda = 7.3$ μm which results in reflection from the array < 7%. In this case, both resonances are in phase, and near-zero reflection from the array is caused by the predominantly forward scattering from the antennas. In turn, a near-zero reflectance means that a generalized Kerker condition is satisfied and hBN antennas effectively serve as Huygens' elements. This effect is complementary to resonant lattice Kerker effect demonstrated earlier for silicon particle arrays [26,27].

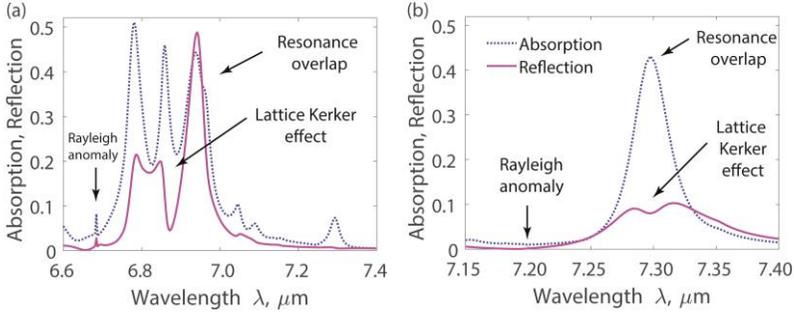

Figure 3. Comparison of reflection and absorption spectra for periods (a) $d_y = 6.7$ μm and (b) $d_y = 7.2$ μm. Profiles are shown for the wavelength larger than Rayleigh anomaly: (a) $\lambda_{RA} = 6.7$ μm and (b) $\lambda_{RA} = 7.2$ μm. Both cases correspond to the overlap of lattice resonance and unaffected antenna resonance, and the overlap results in a significant suppression of reflection. The legend is the same for both panels. Hexagonal boron nitride resonator dimensions are $a_x = 0.9$ μm, $a_y = 0.5$ μm, and $a_z = 2$ μm; array period in $x$-direction $d_x = 2.5$ μm; and the electric field $E$ of the incident light is along $y$-direction.

## CONCLUSION

To conclude, we have studied an array of hBN particles in a spectral range where hBN experiences hyperbolic dispersion and exhibit properties similar to Type II hyperbolic metamaterials. Hexagonal boron nitride particles support multipole resonances there, and higher-order multipoles appear at the larger wavelength with respect to dipole ones. These antennas support excitations of magnetic resonances which brings an additional degree of freedom in designing metasurfaces with subwavelength scatterers. We have considered arrays with different periods and Rayleigh anomaly in the proximity of particle multipole resonances. We have theoretically shown the excitations of lattice resonances that closely follow Rayleigh anomaly. Because of the polarization dependence, in our case, lattice resonances can be excited only for magnetic dipole, electric quadrupole, and higher-order moments, but not for electric dipoles and magnetic quadrupoles. Upon overlap of the antenna multipole resonances, we have observed a decrease of reflection with resonant scattering forward by the antennas in the array. It corresponds to resonant lattice Kerker effect with a generalized condition satisfied by scattering of different multipoles.

The possibility to control resonance positions by the lattice dimensions and reflection suppression provides an opportunity for further development of ultra-thin optical elements, metasurfaces, and improved efficiency of photonic devices. In the present work, we have not performed optimization and theoretically demonstrated a design with scattering elements about 2-μm-thick, which is comparable with dimensions of current state-of-the-art silicon optical elements and metasurfaces in the mid-infrared wavelength range. However, the hyperbolic medium supports high-k modes, an effective mode index can be higher than the one in high-index dielectric structures, and the thickness of hBN antennas with strong resonances can be further decreased down to several hundreds of nanometers (see multipole mode analysis in [3]). For the medium with hyperbolic dispersion, anomalous scaling law of resonant wavelength vs. antennas size [50] enables achieving strong resonances with small dimensions. It means that the radiative optical losses are relatively low, and the remaining limiting factor for further antenna decrease is mainly non-radiative losses.

ACKNOWLEDGMENT

This material is based upon work supported by the Air Force Office of Scientific Research under Grant No. FA9550-16-1-0088.